\documentclass[aps,prl,twocolumn,showpacs,showkey,square,numbers,amssymb,amsmath,nobibnotes,footinbib]{revtex4}
\usepackage{bm}
\usepackage{amsmath}
\usepackage{amsfonts}
\usepackage{amssymb}   
\usepackage{verbatim}
\usepackage{times}
\usepackage{graphicx}
\usepackage{dsfont}
\usepackage{titletoc}

\DeclareBoldMathCommand\boldlangle{\left\langle}
\DeclareBoldMathCommand\boldrangle{\right\rangle}

\newcommand{\barr}{\begin{eqnarray}}
\newcommand{\earr}{\end{eqnarray}}

\newcommand{\beq}{\begin{equation}}
\newcommand{\eeq}{\end{equation}}
\newcommand{\be}{\begin{equation}}
\newcommand{\ee}{\end{equation}}

\newcommand{\dd}{\mathrm{d}}

\DeclareMathOperator\arctanh{arctanh}
\DeclareMathOperator\arccoth{arccoth}
\DeclareMathOperator\arcsinh{arcsinh}

\newcommand{\beeq}[1] {\begin{equation}\begin{split}#1\end{split}\end{equation}}
\newcommand{\beeqn}[1] {\begin{equation*}\begin{split}#1\end{split}\end{equation*}}

\begin{document}

\title{Phase transitions in the condition number distribution of Gaussian random matrices}

\author{Isaac P\'erez Castillo$^1$, Eytan Katzav$^2$ and Pierpaolo Vivo$^3$}
\affiliation{1. Departamento de Sistemas Complejos, Instituto de F\'isica, UNAM, P.O. Box 20-364, 01000 M\'exico D.F., M\'exico\\
2. Department of Mathematics, King's College London, Strand, London WC2R 2LS, United Kingdom\\
$3.$ Laboratoire de Physique Th\'{e}orique et Mod\`{e}les
Statistiques (UMR 8626 du CNRS), Universit\'{e} Paris-Sud,
B\^{a}timent 100, 91405 Orsay Cedex, France}

\date{\today}

\begin{abstract}
We study the statistics of the condition number $\kappa=\lambda_{\mathrm{max}}/\lambda_{\mathrm{min}}$ (the ratio between largest and smallest squared singular values) of $N\times M$ Gaussian random matrices.
Using a Coulomb fluid technique, we derive analytically and for large $N$ the cumulative $\mathcal{P}[\kappa<x]$ and tail-cumulative $\mathcal{P}[\kappa>x]$ distributions of $\kappa$. We find that these distributions decay as $\mathcal{P}[\kappa<x]\approx\exp\left(-\beta N^2 \Phi_{-}(x)\right)$ and $\mathcal{P}[\kappa>x]\approx\exp\left(-\beta N \Phi_{+}(x)\right)$, where $\beta$ is the Dyson index of the ensemble. The left and right rate functions $\Phi_{\pm}(x)$ are independent of $\beta$ and calculated exactly for any choice of the rectangularity parameter $\alpha=M/N-1>0$. Interestingly, they show a weak non-analytic behavior at their minimum $\langle\kappa\rangle$ (corresponding to the average condition number), a direct consequence of a phase transition in the associated Coulomb fluid problem. Matching the behavior of the rate functions around $\langle\kappa\rangle$, we determine exactly the scale of typical fluctuations $\sim\mathcal{O}(N^{-2/3})$ and the tails of the limiting distribution of $\kappa$. The analytical results are in excellent agreement with numerical simulations. 
\end{abstract}

\pacs{5.40.-a, 02.10.Yn, 02.50.Sk, 24.60.-k}


\maketitle

{\it Introduction -} A classical task in numerical analysis is to find the solution $\mathbf{x}$ of a linear system $A\mathbf{x}=\mathbf{y}$, where in the simplest setting $A$ is a square $N\times N$ matrix and $\mathbf{y}$ is a given column vector. The solution can be formally written as $\mathbf{x}=A^{-1}\mathbf{y}$ provided that $A$ is invertible. A very important issue for numerical stability is: how does a small change in the entries of $\mathbf{y}$ or of $A$ propagate on the solution $\mathbf{x}$?\\
The system of equations above is said to be \emph{well}- (or \emph{ill}-) \emph{conditioned} if a small change in the coefficient matrix $A$ or in the right hand side $\mathbf{y}$ results in a small (or large) change in the solution vector $\mathbf{x}$. An ill-conditioned system produces a solution that cannot be trusted, as numerical inaccuracies in the inputs are amplified and propagated to the output \cite{smale}. \\
A standard indicator of the reliability of numerical solutions is the \emph{condition number} (CN) $\kappa=\lambda_{\mathrm{max}}/\lambda_{\mathrm{min}}\geq 1$, where $\lambda_{\mathrm{min}}$ and $\lambda_{\mathrm{max}}$ are the smallest and largest \emph{squared} singular values of $A$, i.e. the positive eigenvalues of $A A^T$ (the square root $\sqrt{\kappa}$ of CN is alternatively used very frequently). The quantity $\ln_b\kappa$ is essentially a worst-case estimate of how many base-$b$ digits are lost in solving numerically that linear system, which is \emph{singular} if $\kappa$ is infinite, \emph{ill-conditioned} if $\kappa$ is ``too large", and \emph{well-conditioned} if $\kappa$ is close to its minimum value $1$.\\ 
Computing $\kappa$ for a large coefficient matrix $A$ in a fast and efficient way can be, however, as difficult a task as solving the original system in the first place \cite{avron}. To overcome this problem, Goldstine and von Neumann \cite{neumann,neumann2} proposed instead to study the generic features of $\kappa$ associated to a \emph{random} matrix $A$ with normally distributed elements \cite{demmel}. What is the \emph{typical} (expected) CN for a system of size $N$? And what is a sensible estimate for the size of its fluctuations?\\
Modern applications of a random condition number of more general (rectangular) matrices $N\times M$ include wireless communication systems \cite{e1,e2,e3,e4,e5}, spectrum sensing algorithms \cite{e6,e7,e8}, convergence rate of iterative schemes \cite{strang}, compressed sensing \cite{sensing}, finance \cite{finance}, meteorology \cite{meteo} and performance assessment of principal component analysis \cite{PCA} among others.\\
The statistics of $\sqrt{\kappa}$ was first computed by Edelman \cite{edel1} for $2\times M$ random Gaussian matrices, as well as the limiting distribution of $\sqrt{\kappa}/N$ for large $N\times N$ matrices. The rectangular case was recently considered in \cite{jiang}. Different bounds for the tails were given in \cite{edel2,chen,azais}. Exact formulae for the distribution of $\kappa$ for finite $N,M$ also exist in terms of cumbersome series of zonal polynomials \cite{anderson,rat} or an integral of a determinant \cite{matt}, whose evaluation becomes impractical even for moderate matrix sizes. Approximate results for correlated non-central Gaussian matrices can be found e.g. in \cite{wei}. Other definitions for the CN also exist \cite{demmel,edel3}.\\
Unfortunately, almost nothing is known about the most dreaded (or welcomed) scenarios for applications, namely the occurrence of \emph{atypical} instances, where the CN is much larger (or smaller) than its expected value. In this Letter, by suitably adapting the Coulomb fluid method of statistical mechanics, we provide an analytical solution to this outstanding problem for large rectangular instances. We show that the large deviation statistics of the CN of Gaussian matrices, expressed in terms of elementary functions, has a rich and elegant structure. As a bonus, we also derive the scale of typical fluctuation of the CN around $\langle\kappa\rangle$, and the tails of its limiting distribution. Let us first summarize our setting and main results.

\begin{figure}
\centering
\includegraphics[width=0.5\textwidth]{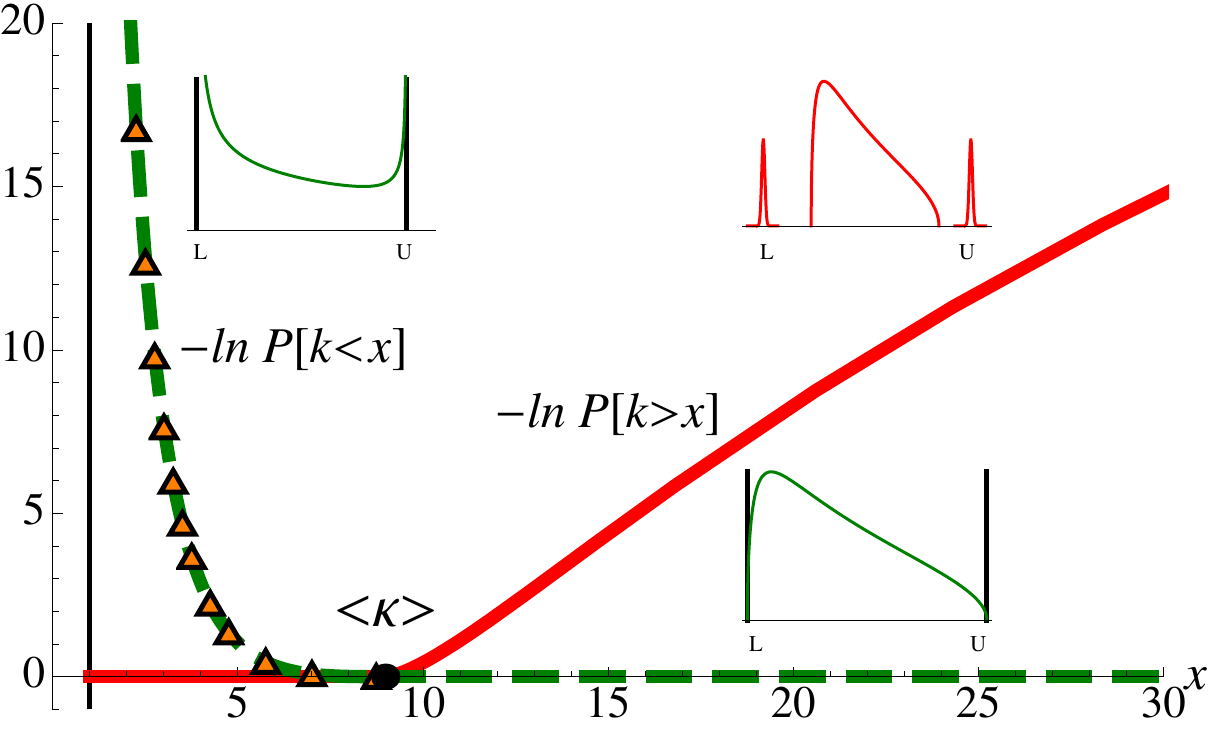} 
\caption{Plot of $-\ln\mathcal{P}[\kappa<x]$ (dashed green, Eq. \eqref{summary1}) and $-\ln\mathcal{P}[\kappa>x]$ (solid red, Eq. \eqref{summary2}), together with numerical simulations for the left branch \cite{supp}. The two rate functions $\beta N^2 \Phi_-(x)$ and $\beta N\Phi_+(x)$ freeze to the zero value upon crossing $\langle\kappa\rangle$. The insets describe the corresponding phases of the Coulomb fluid (active \emph{vs.} inactive barriers for $\Phi_-(x)$ and the pulling of individual extreme charges for $\Phi_+(x)$). An arbitrary value of $N$ has been chosen to produce a  reasonably looking plot in which the two branches are visible.}
\label{fig:rate}
\end{figure}

{\it Summary of results -} We consider rectangular $N\times M$ ($M>N$) matrices $A$ with Gaussian distributed entries (real, complex or quaternions, labelled by the Dyson index $\beta=1,2,4$ respectively, or actually for general $\beta>0$ as discussed in \cite{dumitriu}). Forming the corresponding $N\times N$ covariance matrix $W=A A^T$ \footnote{Here $^T$ stands for transpose ($\beta=1$), hermitian conjugate ($\beta=2$) and symplectic conjugate ($\beta=4$).}, which defines the Wishart ensemble \cite{wishart}, we define its rectangularity parameter $\alpha=M/N-1>0$ and the CN $\kappa=\lambda_{\mathrm{max}}/\lambda_{\mathrm{min}}>1$. Here $\lambda_{\mathrm{max}}$ and $\lambda_{\mathrm{min}}$ are the largest and smallest eigenvalues of $W$. We consider the cumulative $\mathcal{P}[\kappa<x]$ and tail-cumulative (also known as exceedance or survival function) $\mathcal{P}[\kappa>x]$ distributions of $\kappa$, when $N$ and $M$ are large and $\alpha$ is kept finite. Using a Coulomb fluid technique we find that for large $N$ both distributions obey large deviation laws, namely they decay for large $N$ as \footnote{Here $\approx$ stands for the logarithmic equivalence $\lim_{N\to\infty} -\ln \mathcal{P}[\kappa<x]/\beta N^2=\Phi_-(x)$ and similarly for the tail-cumulative branch.}
\begin{align}
\mathcal{P}[\kappa<x] &\approx\exp\left(-\beta N^2 \Phi_{-}(x)\right),\label{summary1}\\
\mathcal{P}[\kappa>x] &\approx\exp\left(-\beta N \Phi_{+}(x)\right)\label{summary2}.
\end{align}
The left and right rate functions $\Phi_\pm (x)$ (depending parametrically on $\alpha$, but not on $\beta$) are given in \eqref{rateleft} and \eqref{rateright} and plotted in Fig. \ref{fig:rate}. Both functions are supported on $x\in (1,\infty)$ and have a minimum (zero) at $\langle\kappa\rangle=[(1+\sqrt{1+\alpha})/(1-\sqrt{1+\alpha})]^2>1$. Therefore the corresponding \emph{density} of $\kappa$ is peaked around $\langle\kappa\rangle$, which is precisely its mean value for large $N$ \footnote{The typical value $\langle\kappa\rangle$ for large $N$ is just the ratio of the average $\langle\lambda_{\mathrm{max}}\rangle=(1+\sqrt{1+\alpha})^2$ and the average $\langle\lambda_{\mathrm{min}}\rangle=(1-\sqrt{1+\alpha})^2$}. Crossing $\langle\kappa\rangle$, both functions freeze to the zero value, and around $\langle\kappa\rangle$ they have an interesting non-analytic behavior, characterized by a third-order (for $\Phi_-(x)$) \cite{majumdar_schehr_13} and second-order (for $\Phi_+(x)$) discontinuity. Both these non-analytic behaviors and the different scaling with $N$ between \eqref{summary1} and \eqref{summary2} are direct consequences of \emph{freezing phase transitions} \cite{texier} in an associated Coulomb fluid problem. The physics of the two branches is however entirely different (see below for details). Matching the behavior of the rate functions around $\langle\kappa\rangle$, we also determine exactly the size ($\sim\mathcal{O}(N^{-2/3})$) of typical fluctuations of $\kappa$ and the tails of its limiting distribution. We now begin by recalling some well-known facts about Wishart matrices.

\begin{figure}
\centering
\includegraphics[width=0.4\textwidth]{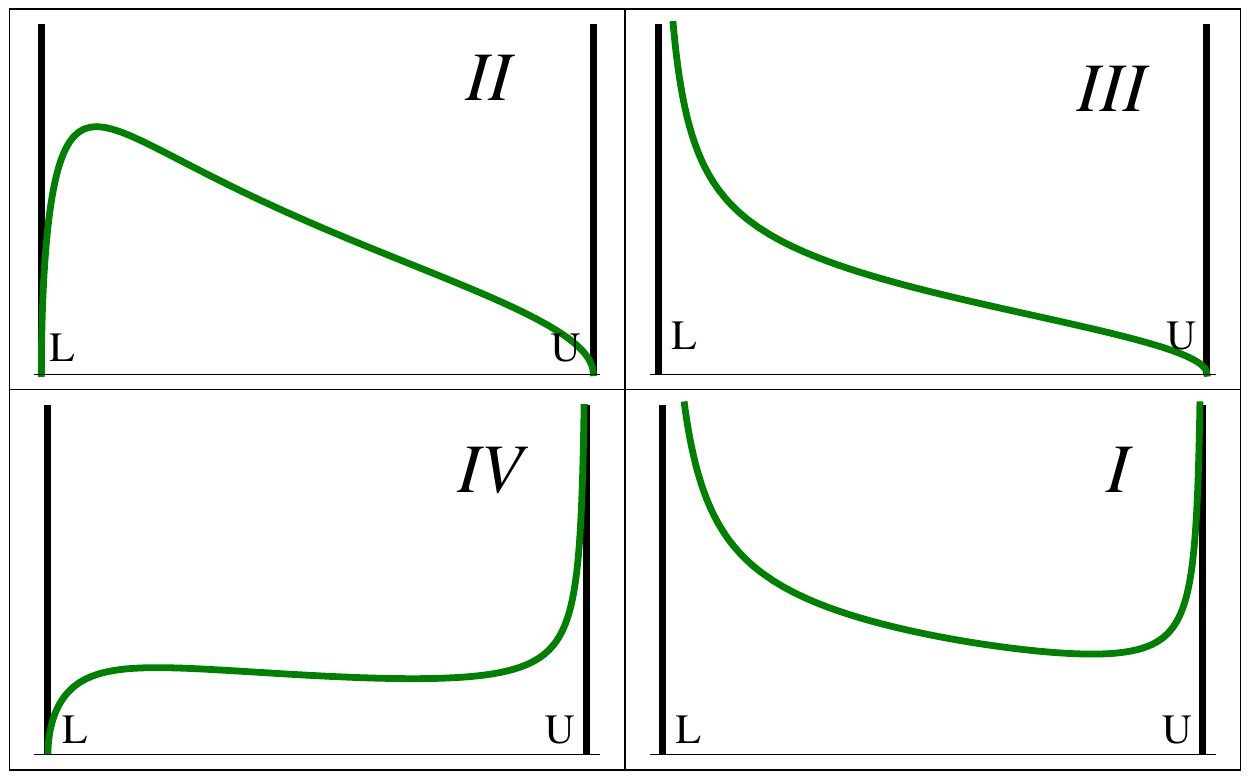}\\
\vspace{.5cm}
\includegraphics[width=0.4\textwidth]{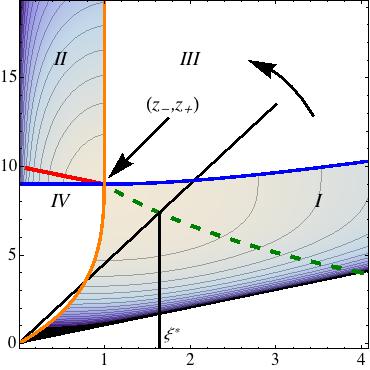} 
\caption{Top: The four phases of the fluid. Region I: the two barriers compress effectively the fluid. Region II: MP law, where the barriers do not affect the fluid. Region III and IV: only the lower or the upper barrier is active, respectively (this scenario is not realized in our CN setting). The analytical expressions in regions $\mathrm{III}$ and $\mathrm{IV}$ where first derived in \cite{katzav_perezCastillo_10,vivo_majumdar_bohigas_07}, respectively.\\
 Down: Regions in the $(L,U)$ plane where the density Eq. \eqref{eqdensity} has different shapes, according to top panels, for $\alpha=3$. We plot level curves of the action \eqref{action} $S[\rho^\star,C;L,U]$ ($\mathrm{I}$) and energy difference $\Delta e(L,U)$ [Eq. \eqref{deltaesupp} in \cite{supp}] ($\mathrm{IV}$). On the dashed green and solid red extremal lines, the action and the energy difference are minimal, respectively. The abscissas $\xi^\star$ (solution of the saddle point equations in the two cases) are given by the intersection of the straight line $U=xL$ (solid black, with the left arrow pointing in the direction of increasing slope $x$) with such extremal lines. Besides, the solid orange line corresponds to the condition $x_{-}{(L,U)}=L$ along which the lower barrier is ineffective (and the upper is effective), while the solid blue line correponds to the condition $x_{+}(L,U)=U$ along which the upper barrier is ineffective (and the lower is effective).}
\label{fig:pic2}
\end{figure}

{\it Generalities -}
The joint probability density (jpd) of the $N$ (real and positive) eigenvalues is given by \cite{fisher,hsu}
\be
P_\beta(\bm{\lambda})=\frac{1}{Z_{0}}\mathrm{e}^{-\frac{1}{2}\sum_{i=1}^ N\lambda_{i}}\prod_{i=1}^N \lambda_i^{\frac{\beta}{2}(\alpha N+1)-1}\prod_{i<j}|\lambda_i-\lambda_j|^{\beta},\label{jpd}
\ee
where $Z_0$ is a normalization constant and $\beta=1,2,4$ is the Dyson index of the ensemble.  Balancing the first and third terms in \eqref{jpd}, it is quite easy to estimate that the typical scale of an eigenvalue is $\sim\mathcal{O}(N)$. Thus, after rescaling $\lambda_i\to\beta N\lambda_i$, the jpd \eqref{jpd} can be rewritten in the form $P_\beta(\bm{\lambda})\propto \exp(-\beta N^2 E[\{\bm{\lambda}\}])$, where the $\mathcal{O}(1)$ \emph{energy}  is
\be
E[\{\bm{\lambda}\}]=\frac{1}{2N}\sum_{j=1}^N\lambda_j-\frac{\alpha}{2N}\sum_{j=1}^N\ln\lambda_j-\frac{1}{2N^2}\sum_{j\neq k}\ln|\lambda_j-\lambda_k|.\label{energy}
\ee
Written in this form, the jpd \eqref{jpd} resembles the Gibbs-Boltzmann canonical weight of a 2D fluid of charged particles, confined on the semi-infinite (positive) line and in equilibrium at inverse temperature $\beta$ under competing interactions: the external linear-logarithmic potential in \eqref{energy} drives the charges towards its minimum, while the third term (representing an all-to-all repulsive interaction of the Coulomb type in 2D) spreads them apart. This thermodynamical analogy, originally pioneered by Dyson \cite{dyson_62}, has been lately employed in several different contexts \cite{dean_majumdar_06,dean_majumdar_08,majumdar_schehr_13,vivo_majumdar_bohigas_07, majumdar_vergassola_09, katzav_perezCastillo_10}. \\
The average spectral density of the Wishart model $\rho(\lambda)=N^{-1}\sum_{i=1}^N\boldlangle\delta(\lambda-\lambda_i)\boldrangle$ (where $\boldlangle\cdots\boldrangle$ denotes averaging with respect to the jpd \eqref{jpd}) is expected for large $N$ to have the scaling form $\rho(\lambda)=N^{-1}\rho_{\mathrm{mp}}(\lambda/N)$, where the function $\rho_{\mathrm{mp}}(x)=\frac{1}{2\pi x}\sqrt{(x-z_-)(z_+-x)}$
is the celebrated Mar\v{c}enko-Pastur (MP) law on the compact support (for $\alpha>0$) $x\in [z_-,z_+]$ with $z_\pm=(1\pm \sqrt{\alpha+1})^2$. This MP law is a particular case of the general solution \eqref{eqdensity} of the integral equation \eqref{loginteq} below (see Fig. \ref{fig:pic2}, top) when the two barriers $L,U$ are ineffective ($L\leq z_-$ and $U\geq z_+$). We start now by considering the cumulative distribution of the CN first, and get to the tail-cumulative afterwards. 

{\it Cumulative distribution -} The cumulative distribution $\mathcal{P}[\kappa<x]$ of the CN $\kappa$ (depending parametrically on $\beta$ and $\alpha=M/N-1>0$), can be written as \cite{matt,supp}
\be
\mathcal{P}[\kappa<x]=\frac{1}{(N-1)!}\int_0^\infty \hspace{-1mm}\dd\lambda_1 \left[\int\cdot\cdot\int_{\lambda_1}^{x\lambda_1} \prod_{j=2}^N\dd\lambda_j\ P_\beta(\bm{\lambda})\right].\label{Pk}
\ee
The goal is to evaluate this multiple integral for large $N$ by the Coulomb fluid method. The first step is to rewrite the jpd \eqref{jpd} in the Gibbs-Boltzmann form described above. Here, $N-1$ fluid particles are however not free to spread on the whole positive line, but instead \emph{constrained} to live within the box $[\lambda_1,x\lambda_1]$, where $\lambda_1$ is the (free) position of the leftmost particle.\\
The second step consists of a coarse-graining procedure, where one introduces a normalized density of particles $\rho(\lambda)=(N-1)^{-1}\sum_{i=2}^N\delta(\lambda-\lambda_i)$ for the $N-1$ particles $\lambda_i$ $(i\neq 1)$ living \emph{inside} the box. Using the replacement rule $\sum_{i>1} g(\lambda_i)=(N-1)\int\dd\lambda\rho(\lambda)g(\lambda)$, we can convert the energy function $E[\{\bm{\lambda}\}]$ into a continuous action $S$ (depending on $\rho$, and parametrically on the location of the leftmost particle $\lambda_1$ and $x$). The multiple integration \eqref{Pk} is therefore interpreted as the canonical partition function of the associated Coulomb fluid, where the sum over all microscopic configurations of $\{\bm\lambda\}$ compatible with the normalized density $\rho$ amounts to a functional integration over $\rho$ and a standard integration over $\lambda_1$. Eventually, these resulting integrals are evaluated using the saddle point method. Performing these steps, we get 
\be
\mathcal{P}[\kappa<x]\propto\int_0^\infty\hspace{-2mm}\dd\xi\int\mathcal{D}[\rho,C]\mathrm{e}^{-\beta N^2 S[\rho,C;\xi,x\xi]},\label{pbulkN}
\ee
where we renamed $\lambda_1\to\xi$ for later convenience and the action of this fluid (confined between the lower $L$ and upper $U$ barriers of the box) is 
\begin{align}
\nonumber S[\rho,C;L,U] &=\int_L^U\dd\lambda\rho(\lambda) V(\lambda)+C\\
&-\frac{1}{2}\iint_L^U\dd\lambda\dd\lambda^\prime\rho(\lambda)\rho(\lambda^\prime)\ln|\lambda-\lambda^\prime|.
\label{action}
\end{align}
Here $V(\lambda)=(\lambda-\alpha\ln\lambda)/2-C$ and $C$ is a Lagrange multiplier enforcing normalization of $\rho$. Eq. \eqref{action} is easily identified as the continuous version of the energy Eq. \eqref{energy}, where we have neglected subleading $\mathcal{O}(N)$ contributions \cite{supp}. \\
Evaluating the functional integral in \eqref{pbulkN} by the saddle point method, $\frac{\delta S}{\delta\rho}\Big|_{\rho=\rho^\star}=0$, we get
\be
 V(\lambda)=\int_{L}^U\dd\lambda^\prime\rho^\star(\lambda^\prime)\ln|\lambda-\lambda^\prime|,\label{loginteq}
\ee
where the solution $\rho^\star(\lambda)$ is just the equilibrium density of the Coulomb fluid constrained to live within the box $[L,U]$. Clearly, if we release the barriers $L,U$ we expect to recover the unconstrained MP law, $\rho^\star(\lambda)\to \rho_{\mathrm{mp}}(\lambda)$. 
\noindent Solving this integral equation for a normalized $\rho^\star$ between two barriers at $L$ and $U$ is one of the main technical challenges that we managed to overcome. Skipping details \cite{supp}, we find that the general solution of \eqref{loginteq} is \cite{tricomi}
\be
\rho^\star(\lambda)=\frac{\left(x_+(L,U)-\lambda\right)\left(\lambda-x_{-}(L,U)\right)}{2\pi\lambda\sqrt{(U-\lambda)(\lambda-L)}}\mathds{1}_{[L,U]}(\lambda),\label{eqdensity}
\ee
where $x_+(L,U)\geq U>L\geq x_-(L,U)$,  $x_{\pm}(L,U)$ are the roots of
$
x^2-x\left(\frac{L+U}{2}+\alpha+2\right)+\alpha\sqrt{LU}=0,
$
and $\mathds{1}_{[a,b]}(x)=1$ if $x\in [a,b]$ and $0$ otherwise.\\
How does this density look like for given values of $L$ and $U$? Four different shapes (phases of the fluid) are possible \cite{supp} for $\alpha>0$ that are plotted in Fig. \ref{fig:pic2} (top). For example, setting $(L,U)=(z_-,z_+)$ the corresponding density \eqref{eqdensity} is the MP law $\rho^\star(x)=\rho_{\mathrm{mp}}(x)$ (phase $\mathrm{II}$). This critical MP point, which is  marked in Fig. \ref{fig:pic2} (bottom),  separates region $\mathrm{II}$, where the barriers are \emph{ineffective} ($L<z_-$ and $U>z_+$) and the equilibrium density is again just $\rho_{\mathrm{mp}}(x)$, from region $\mathrm{I}$ (where the barriers are instead \emph{effective} in compressing the MP sea).\\
Once we have evaluated the functional integral by the saddle-point method (which implies inserting the density \eqref{eqdensity} in the action \eqref{action}) we set $L=\xi$ and $U=x\xi$ and evaluate the remaining $\xi$-integral again by the saddle-point method. This yields an optimal value $\xi^\star$ as the solution of $\frac{\dd}{\dd\xi}S[\rho^\star,C;\xi,x\xi]\Big|_{\xi=\xi^\star}=0$. This value $\xi^\star(x)$ is marked in Fig. \ref{fig:pic2} (bottom) as the intersection between the straight line $U=xL$ of varying slope $x>1$ and the dashed green line on which the action $S[\rho^\star,C;L,U]$ is minimal.\\   
The final result reads
$
\mathcal{P}[\kappa<x]\approx\mathrm{e}^{-\beta N^2 \Phi_-(x)}
$, where the $\mathcal{O}(N^2)$ decay is traced back to the high energy cost in compressing the whole sea of strongly correlated particles.
Here $\Phi_-(x)=S[\rho^\star,C;\xi^\star,x\xi^\star]-S[\rho_{{\rm mp}},C;z_-,z_+]$, where the second term comes from the normalization factor and needs to be subtracted. We eventually obtain
\be
\Phi_-(x)=
\frac{1}{8}\left[f_1^{(\alpha)}(1+\sqrt{x})+\ln f_2^{(\alpha)}(1+\sqrt{x})\right]\mathds{1}_{(1,\langle\kappa\rangle)}(x),\label{rateleft}
\ee
where $f_{1,2}^{(\alpha)}(\omega)$ are elementary functions listed in \cite{supp}. The rate function $\Phi_-(x)$ thus freezes to the value $0$ as $x$ increases up to the critical value $\langle\kappa\rangle$ (implying $\xi^\star(\langle\kappa\rangle)\to z_-$). Beyond this limit, the barriers are no longer effective, and new physical insights are needed to tackle the tail-cumulative regime (see next subsection). The limits are $\Phi_-(x\to\langle\kappa\rangle^{-})\sim K(\alpha)(\langle\kappa\rangle-x)^{3}$ and $\Phi_-(x\to 1^{+})\sim (-1/2)\ln(x-1)$, where $K(\alpha)= -(-1+\sqrt{1+\alpha})^8/96\sqrt{1+\alpha}(1+\sqrt{1+\alpha})^4$. This implies a third order discontinuity across $\langle\kappa\rangle$ as anticipated. Also, close to $1$, the density of $\kappa$ has the following power-law tail,  $\mathcal{P}[\kappa = 1+\epsilon]\sim \epsilon^{\beta N^2/2}$ to leading order in $N$ for $\epsilon\to 0^+$. Although formally valid only for $\alpha>0$, it turns out that in the limit $\alpha\to 0$ (square Gaussian matrices, where the scaling with $N$ is different) the rate function \eqref{rateleft} is still well-behaved and we recover Edelman's result \cite{edel1} to leading order in $N$ \cite{supp}. We now turn to the tail-cumulative distribution (the right branch in Fig. \ref{fig:rate}).

{\it Tail-cumulative distribution -} Contrary to the previous case, the tail-cumulative distribution $\mathcal{P}[\kappa>x]$ does \emph{not} admit a multiple-integral representation of the type \eqref{Pk}, which could be mapped to the physics of a fluid ``trapped" between two hard barriers.\\
The starting point of the calculation is again the energy function \eqref{energy}, though. The Coulomb fluid physics suggests that atypically large values of the CN $\kappa=\lambda_{\mathrm{max}}/\lambda_{\mathrm{min}}$ are obtained when the right-and-leftmost particles are \emph{pulled} away from the MP sea in opposite directions $(\lambda_{\mathrm{max}}-\lambda_{\mathrm{min}}\sim\mathcal{O}(N))$, a procedure that is energetically not able to generate macroscopic rearrangements \emph{within} the MP sea. This elegant energetic argument was first introduced in \cite{majumdar_vergassola_09}. Following this physical picture, the right rate function $\Phi_+(x)$ is determined by the $\mathcal{O}(N)$ energy cost $\Delta E(L,U)$ in pinning the leftmost and rightmost charges at $L$ and $U$, well outside the \emph{unperturbed} MP sea in between (see red inset in Fig. \ref{fig:rate}). The level curves of the $\mathcal{O}(1)$ $\Delta e(L,U):=\Delta E(L,U)/\beta N$ are depicted in region $\mathrm{II}$ of Fig. \ref{fig:pic2}, together with the extremal line (solid red) where it attains its minimum value. Setting now $L=\xi$ and $U=\xi x$, the energetically most favored position $\xi^\star$ for the leftmost outlier will be determined again by the intersection point of that extremal line and the straight line $U=xL$.\\
Skipping details \cite{supp}, this change in energy can be written for large $N$ as 
\begin{align}
\nonumber\Delta e(\xi,\xi x) &=\frac{(\xi-z_-)+(\xi x-z_+)}{2}-\frac{\alpha}{2}\ln\frac{\xi^2 x}{z_- z_+}\\
&-\int_{z_-}^{z_+}\dd\eta \rho_{\mathrm{mp}}(\eta)\ln\left|\frac{(\xi-\eta)(\eta-\xi x)}{(z_- - \eta)(\eta-z_+)}\right|.
\label{deltae1}
\end{align}
The probability of this ``pinned" configuration of eigenvalues (yielding a CN $\kappa$ \emph{exactly} equal to $x$) is $\propto\exp(-\beta N\Delta e(\xi,\xi x))$. Finding the optimal position $\xi^\star$ for the leftmost particle by minimizing \eqref{deltae1}  with respect to $\xi$, we eventually obtain $\mathcal{P}[\kappa>x] \approx\exp\left(-\beta N \Phi_{+}(x)\right)$, where $\Phi_{+}(x)=\Delta e(\xi^\star(x),\xi^\star(x) x)$ is given by \footnote{Note that this energetic argument gives (strictly speaking) the \emph{density} of $\kappa$, $\mathcal{P}[\kappa=x]$ and not its tail-cumulative distribution $\mathcal{P}[\kappa>x]$. However, the large $N$ decay of the two is the same to leading order.}
\begin{equation}
\Phi_+(x)=\ln\left[\left(g^{(\alpha)}(x)\right)^{\alpha/2}\left(h^{(\alpha)}(x)\right)^{2(2+\alpha)}\right]\mathds{1}_{(\langle\kappa\rangle,\infty)}(x).\label{rateright}
\end{equation}
The functions $g^{(\alpha)}(x)$ and $h^{(\alpha)}(x)$ have lengthy but explicit expressions in terms of elementary functions \cite{supp}. The rate function $\Phi_+(x)$ again freezes to the value $0$ as $x$ decreases down to the critical value $\langle\kappa\rangle$ (implying $\xi^\star(\langle\kappa\rangle)\to z_-$), where the pinned outliers reconnect with the MP sea. The limits are $\Phi_+(x\to\langle\kappa\rangle^{+})\sim J(\alpha)(x-\langle\kappa\rangle)^{3/2}$ and $\Phi_+(x\to \infty)\sim (\alpha/2)\ln x$, where $J(\alpha)=\sqrt{2} \sqrt[4]{\alpha+1} \left(\sqrt{\alpha+1}-1\right)^4/3 \sqrt{\alpha+2} \left(\sqrt{\alpha+1}+1\right)^2$. This implies a second order discontinuity across $\langle\kappa\rangle$ as anticipated. Also, at infinity the density of $\kappa$ therefore decays as a power-law,  $\mathcal{P}[\kappa = x]\sim x^{-\alpha\beta N/2}$ to leading order in $N$ for $x\to \infty$.

{\it Conclusions -} In summary, we have computed analytically for large $N$ the cumulative and tail-cumulative distributions of the CN $\kappa$ of rectangular $N\times M$ Gaussian random matrices. Mapping the problem to the calculation of the free energy of an associated Coulomb fluid, we found that the random variable $\kappa$ satisfies large deviation laws governed by rate functions $\Phi_\pm(x)$ to the left and to the right of the expected value $\langle\kappa\rangle$, albeit with different scalings with $N$. These rate functions (see \eqref{rateleft} and \eqref{rateright}) essentially describe the probability of sampling a random Gaussian matrix with an atypically large (or small) CN. They are monotonic and convex, with a non-analytic behavior at their zero $\langle\kappa\rangle$. This non-analytic behavior is a direct consequence of a very rich thermodynamics of the associated Coulomb fluid. More precisely, across $\langle\kappa\rangle$ a \emph{compressed} (left branch) or \emph{stretched} (right branch) fluid undergoes \emph{freezing} phase transitions of different orders (third and second respectively). \\
Matching the behavior of the rate functions close to their minimum $\langle\kappa\rangle$, we deduce that \emph{typical} fluctuations of $\kappa$ around $\langle\kappa\rangle$ should occur on a scale of $\mathcal{O}(N^{-2/3})$, and setting $\chi=f(\alpha)N^{2/3}(\kappa-\langle\kappa\rangle)$, with $f(\alpha)=2^{1/3}(1+\alpha)^{1/6}(-1+\sqrt{1+\alpha})^{8/3}/(1+\sqrt{1+\alpha})^{4/3}$, the scaled random variable $\chi$ has a $N$- and $\alpha$-independent distribution $\mathcal{P}[\chi<x]=\mathcal{F}_\beta(x)$ with tails $\sim\exp\left(-\beta |x|^3\right)$ for $x\to -\infty$ and $\sim\exp\left(-\beta x^{3/2}\right)$ for $x\to\infty$. The scaling is in agreement with a recent result \cite{jiang}, valid only for $M\gg N^3$. Our analytical results have been numerically checked with excellent agreement \cite{supp}.

{\it Acknowledgments -} PV acknowledges financial support from Labex-PALM (project RandMat). We thank G. Schehr and F. D. Cunden for helpful discussions. IPC acknowledges hospitality by the LPTMS (Univ. Paris Sud) where this work was completed.

\begingroup
\renewcommand{\addcontentsline}[3]{}
\renewcommand{\section}[2]{}
\subsection{}

\endgroup

\newpage
\setcounter{equation}{0} 
\numberwithin{equation}{subsubsection}

\begin{widetext}

\section{Supplementary Material}
\tableofcontents

\subsection{Derivation of the expression for the action in Eq. \eqref{action}}
Let us start by noticing that $\mathcal{P}[\kappa<x]={\rm Prob}[\lambda_{{\rm max}}< x\lambda_{{\rm min}}]$, which can be worked out as follows
\beeqn{
\mathcal{P}[\kappa<x]&={\rm Prob}(\lambda_{1}\leq\lambda_{2}\leq \cdots\leq\lambda_{N}\leq\kappa\lambda_{1})\\
&=\int_0 ^{\infty}\dd x {\rm Prob}(x\leq\lambda_{2}\leq\lambda_{3}\leq \cdots\leq \lambda_{N-1}\leq\lambda_{N}\leq\kappa x\,, \lambda_1=x)\\
&=\frac{1}{(N-1)!}\int_0^\infty \dd x{\rm Prob}(\lambda_1=x\,,\,\{\lambda_{1}\leq\lambda_{i}\leq \kappa\lambda_1\,,i=2,\ldots,N\})
}
or in terms of the jpd of eigenvalues of the Wishart ensemble we write
\beeqn{
\mathcal{P}[\kappa<x]&=\frac{1}{(N-1)!}\int_0^\infty \dd\lambda_1 \left[\int_{\lambda_1}^{x\lambda_1} \dd \lambda_2\cdots \int_{\lambda_1}^{x\lambda_1}\dd\lambda_{N} P_\beta(\bm{\lambda})\right]
}
To go to a continuous theory we start from Eq. \eqref{jpd} of this Letter, which defines the jpd of eigenvalues of the Wishart ensemble, we first rewrite it in exponential form and rescale all the eigenvalues as $\lambda_i\to\beta N\lambda_i$. This results in $P_\beta(\bm{\lambda})\propto \exp(-\beta N^2 E[\{\bm{\lambda}\}])$, with
\beeqn{
E[\{\bm{\lambda}\}]=\frac{1}{2N}\sum_{j=1}^N\lambda_j-\frac{\alpha}{2N}\sum_{j=1}^N\ln\lambda_j-\frac{1}{2N^2}\sum_{j\neq k}\ln|\lambda_j-\lambda_k|,\label{energysupp}
}
where we have kept only the relevant terms. Next, we rename  $\lambda_1$ as $\xi$ and separate it from the rest of eigenvalues which, abusing notation, are denoted  again as $\bm{\lambda}=(\lambda_2,\ldots,\lambda_N)$, \textit{viz.}
\beeqn{
E[\{\xi,\bm{\lambda}\}]&=\frac{1}{2N}\left(\xi+\sum_{j=2}^N\lambda_j\right)-\frac{\alpha}{2N}\left(\ln\xi+\sum_{j=2}^N\ln\lambda_j\right)-\frac{1}{2N^2}\left(\sum_{j\neq k}\ln|\lambda_j-\lambda_k|+2\sum_{j=2}^N\ln|\xi-\lambda_j|\right)
}
This expression for $E[\{\xi,\bm{\lambda}\}]$ can  now be expressed as a continuous theory in the following form:
\beeq{
S[\rho(\lambda;\bm{\lambda})]\equiv E[\{\xi,\rho(\lambda;\bm{\lambda})\}]&=\frac{1}{2}\int \dd\lambda \rho(\lambda;\bm{\lambda})-\frac{\alpha}{2} \int \dd\lambda \rho(\lambda;\bm{\lambda})\ln\lambda\\
&-\frac{1}{2}\iint \dd\lambda\dd\lambda' \rho(\lambda;\bm{\lambda})\rho(\lambda;\bm{\lambda})\ln|\lambda-\lambda'|+\mathcal{O}(N^{-1})
\label{eq:AS}
}
where  have introduced the normalized one-point function
\begin{align*}
\rho(\lambda;\bm{\lambda}) &=\frac{1}{N-1}\sum_{i=2}^N\delta(\lambda-\lambda_i)\,.
\end{align*}
It is important to note that the dependence on $\xi$ does not appear in the leading terms of Eq. \eqref{eq:AS} and therefore will be irrelevant for large $N$. That is why we write the energy as $S[\rho(\lambda;\bm{\lambda})]$, that is, only dependent on the density $\rho(\lambda)$. Going back to our expression for the cumulative function we write
\beeqn{
\mathcal{P}[\kappa<x]&\propto \int \mathcal{D}[\rho] \int_0^\infty \dd\xi \left[\int_{\lambda_1}^{\kappa\lambda_1} \dd \lambda_2\cdots \int_{\lambda_1}^{\kappa\lambda_1}\dd\lambda_{N} \right] e^{-\beta N^2 S[\rho(\lambda)]}\delta_{(F)}\left(\rho(\lambda) -\frac{1}{N-1}\sum_{i=2}^N\delta(\lambda-\lambda_i)\right)
}
Finally, using a Fourier representation for the functional Dirac delta, integrating its corresponding Lagrange multiplier, and ignoring subleading terms, we end up with the following expression:
\beeqn{
\mathcal{P}[\kappa<x]&\propto \int_0^\infty \dd \xi\int \mathcal{D}[\rho,C]   e^{-\beta N^2 S[\rho,C;\xi,x\xi]}
}
with
\beeq{
 S[\rho,C;L,U] &=\frac{1}{2}\int_L^U\dd\lambda\lambda\rho(\lambda)-\frac{\alpha}{2}\int_L^U\dd\lambda\rho(\lambda)\ln\lambda\\
&-\frac{1}{2}\iint_L^U\dd\lambda\dd\lambda^\prime\rho(\lambda)\rho(\lambda^\prime)\ln|\lambda-\lambda^\prime|-C\left(\int_L^U\dd\lambda\rho(\lambda) -1\right)+\mathcal{O}(N^{-1}).\label{actionsupp2}
}
as in Eq. \eqref{action} of the Letter. 
\subsection{Solution of integral equation Eq.  \eqref{loginteq}}
The integral equation Eq. \eqref{loginteq} of the Letter can be solved resorting to the theory of so-called Carleman equation, which is an integral equation of type
\beeq{
\int_{a}^b\ln|x-t| y(t) \dd t=f(x).\label{carleman}
}
In our case we have $a=L$, $b=U$, $y(t)=\rho^\star(y)$, and $f(x)=\frac{x}{2}-\frac{\alpha}{2}\ln x-C$. For our  purposes, the general solution of \eqref{carleman} reads
\beeqn{
y(x)=\frac{1}{\pi^2\sqrt{(x-a)(b-x)}}\left[\int_{a}^b\frac{\sqrt{(t-a)(b-t)} f'(t) \dd t}{t-x}+B\right].
}
Recalling the following integral results:
\beeqn{
&\int_{a}^b\frac{\sqrt{(t-a)(b-t)} }{t-x}\dd t=\pi\left(\frac{b+a}{2}-x\right)\,,\quad \int_{a}^b\frac{\sqrt{(t-a)(b-t)} }{t(t-x)}\dd t=-\frac{\pi}{x}\left(x-\sqrt{ab}\right)\,,
}
we obtain the following expression for $\rho^{\star}(x)$:
\beeqn{
\rho^{\star}(x)&=\frac{1}{2\pi\sqrt{(x-a)(b-x)}}\Bigg[\frac{b+a}{2}-x+\frac{\alpha}{x}\left(x-\sqrt{ab}\right)+B\Bigg],
}
where $B$ is determined by looking for the physical solution. Imposing normalization and using the following results
\beeqn{
&\int_{a}^{b}\dd x \frac{1}{2\pi\sqrt{(x-a)(b-x)}}=\frac{1}{2}\,,\quad\quad \int_{a}^{b}\dd x \frac{1}{2\pi x\sqrt{(x-a)(b-x)}}=\frac{1}{2\sqrt{ab}},\\
&\int_{a}^{b}\dd x \frac{x}{2\pi \sqrt{(x-a)(b-x)}}=\frac{a+b}{4},
}
 we obtain that $B=2$. Finally, setting $a=L$ and $b=U$ we write
\beeq{
\rho^\star(\lambda)=\frac{\left(x_+(L,U)-\lambda\right)\left(\lambda-x_{-}(L,U)\right)}{2\pi\lambda\sqrt{(U-\lambda)(\lambda-L)}}\mathds{1}_{[L,U]}(\lambda),\label{eqdensitysupp}
}
where $x_{\pm}(L,U)$ are the roots of the polynomial
\beeqn{
 \lambda^2-\lambda\left(\frac{L+U}{2}+\alpha+2\right)+\alpha\sqrt{LU}=0
}
Writing \eqref{eqdensitysupp} in this way is most convenient as we can see that the physical normalized solution is the one such that $x_{-}(L,U)\leq L<U\leq x_{+}(L,U)$. This implies that the Coulomb fluid has four phases:
\begin{itemize}
\item $x_{+}(L,U)=U$ and $x_{-}(L,U)=U$. In this case we notice that the spectral density becomes the MP law (Region II in Fig. 2):
\beeqn{
\rho^\star(\lambda)&=\frac{\sqrt{(U-\lambda)(\lambda-L)}}{2\pi \lambda}\,,\quad\quad L=z_{-}\,,\quad  U=z_{+}\,.
}
\item $x_{+}(L,U)=U$ and $x_{-}(L,U)<L$. In this case the spectral density becomes (Region III of Fig. 2):
\beeqn{
\rho^\star(\lambda)&=\frac{1}{2\pi}\sqrt{\frac{U-\lambda}{\lambda-L}}\left[\frac{\lambda-\alpha\sqrt{L/U}}{ \lambda}\right]\,,\quad L(U)=\frac{1}{U}\left[\alpha+\sqrt{(\alpha-U)^2-4U}\right]^2\,,\quad U\geq \left(1+\frac{1}{\sqrt{c}}\right)^2\,.
}
Note that in this case we take as free parameter the position of the upper barrier $U$, instead of the lower barrier, as the latter will result into a third degree polynomial to solve.
\item $x_{-}(L,U)=L$ and $x_{+}(L,U)>U$. In this case the spectral density becomes (Region IV of Fig. 2):
\beeqn{
\rho^\star(\lambda)&=\frac{1}{2\pi}\sqrt{\frac{\lambda-L}{U-\lambda}}\left[\frac{\alpha\sqrt{U/L}-\lambda}{\lambda}\right]\,,\quad U(L)=\frac{1}{L}\left[\alpha-\sqrt{(\alpha-L)^2-4L}\right]^2\,,\quad L\leq \left(1-\frac{1}{\sqrt{c}}\right)^2\,,
}
where in this case we take the lower barrier $L$ as our free parameter.
\item In this case we have  $x_{+}(L,U)> U$ and $x_{-}(L,U)< L$ and (Region I of Fig. 2):
\beeqn{
\rho^\star(\lambda)&=\frac{[x_{+}(L,U)-\lambda][\lambda-x_{-}(L,U)]}{2\pi \lambda\sqrt{(U-\lambda)(\lambda-L)}}
}
with $x_{\pm}(L,U)$ are the solutions 
\beeqn{
x_{\pm}(L,U)=\frac{1}{2}\left[\left(\frac{L+U}{2}+\alpha+2\right)\pm\sqrt{\left(\frac{L+U}{2}+\alpha+2\right)^2-4\alpha\sqrt{LU}}\right]
}
\end{itemize}

\subsection{Left rate function}
To derive the left rate function we need to evaluate the action at the solution of the saddle point, that is $S[\rho^{\star},C;L,U]$. It is possible to get rid of the double integral appearing \eqref{actionsupp2} by taking the saddle-point equation, multiplying by $\rho^{\star}(\lambda)$, and integrating $\lambda$ over the interval $[L,U]$. The constant $C$ is then determined taking the value of $\lambda=U$ at the saddle-point equation. All in all, the action at the saddle point takes the following form
\beeqn{
S[\rho^\star,C;L,U]&=\frac{1}{4}\int_{L}^{U} \dd\lambda \rho^{\star}(\lambda)\left(\lambda-\alpha\ln\lambda\right)+\frac{1}{2}C(L,U)
}
with
\beeqn{
C(L,U)=\frac{U}{2} -\frac{\alpha}{2}\ln U-\int_{L}^{U}  \dd \lambda\,\rho^{\star}(\lambda) \ln|U-\lambda|
}
Now we report the result for the following integrals:
\beeqn{
\int_{L}^{U} \dd\lambda \rho^{\star}(\lambda)\lambda&=\frac{1}{4} (L+U) \left(\alpha+\frac{L+U}{2}+2\right)-\frac{1}{2} \alpha \sqrt{L U}-\frac{1}{16} \left(3 L^2+2 L U+3 U^2\right)\\
\int_{L}^{U} \dd\lambda \rho^{\star}(\lambda)\ln \lambda&=\frac{1}{4} \left(8 (\alpha+1) \ln \left(\sqrt{L}+\sqrt{U}\right)-2 \alpha \ln (L)-2 \alpha \ln (U)-8 (\alpha+1) \ln (2)+2 \sqrt{L U}-L-U\right)\\
\int_{L}^{U}  \dd\lambda \rho^{\star}(\lambda) \ln|U-\lambda|&=\frac{1}{4} \left(-2 \alpha \ln \left(\frac{U \left(-2 \sqrt{L U}+L+U\right)}{U-L}\right)+2 (\alpha+2) \ln (U-L)-4 (\alpha+2) \ln (2)-L+U\right)
}
Gathering all these results and after simplifying we obtain the final expression for the action
\beeqn{
S[\rho^\star,C;L,U]&=-\frac{1}{64} \Bigg(-2 L (4 \alpha+U+8)+16 \alpha \sqrt{L U}-8 \alpha \left(-4 (\alpha+2) \ln \left(\sqrt{L}+\sqrt{U}\right)+\alpha \ln (L)+\alpha \ln (U)\right)\\
&-8 (\alpha+2) U-32 (\alpha (\alpha+2)+2) \ln (2)+L^2+32 \ln (U-L)+U^2\Bigg)
}
Two more steps need to be done to get the left rate function: the first one is to take $L=\xi$ and $U=x\xi$ and minimize $S[\rho^\star,C;\xi,x\xi]$ with respect to $\xi$. This gives
\beeqn{
\frac{\dd}{\dd \xi}S[\rho^\star,C;\xi,x\xi]\Big|_{\xi=\xi^\star}=0\Rightarrow \xi^{\star}(x)=4\frac{1+\alpha}{(1+\sqrt{x})^2}\,.
}
The second step is to obtain the normalization factor by taking the limit $x\to\infty$ in the action. This, in the Coulomb fluid picture, corresponds to $x\to\langle \kappa\rangle$. By noting that $\xi^{\star}(\langle \kappa\rangle)=z_{-}$ and that $\rho^\star\to \rho_{{\rm mp}}$, the left rate function $\Phi_{-}(x)=S[\rho^\star,C;\xi^{\star},x\xi^{\star}]-S[\rho_{{\rm mp}},C;z_{-},z_{+}]$ gives
\beeqn{
 \Phi_{-}(x)&=\frac{1}{8 \left(\sqrt{x}+1\right)^2}\Bigg[4 \alpha^2 \left(x \ln (2)+\sqrt{x} \ln (4)-\ln \left(\sqrt{x}+1\right)\right)+\alpha^2 \left(\sqrt{x}+1\right)^2 \ln (x)\\
&-2 \left(x+2 \sqrt{x}\right) \left(2 \alpha^2 \ln \left(\sqrt{x}+1\right)+\ln (\alpha+1)\right)\\
&+4 \alpha^2 \left(\sqrt{x}+1\right)^2 \arccoth(2 \alpha+1)+\alpha^2 \ln (16)+2 (\alpha+1) \left(\alpha x-2 (\alpha+2) \sqrt{x}+\alpha\right)\\
   &-2 \ln (\alpha+1)+8 \left(\sqrt{x}+1\right)^2 \arccoth\left(\sqrt{x}\right)\Bigg]\,.
}
Finally,  after some algebra, $\Phi_{-}(x)$ can be rewritten as reported in Eq. \eqref{rateleft}, and in terms of the functions $f_1^{(\alpha)}(\omega)$ and $f_2^{(\alpha)}(\omega)$ 
\begin{align*}
f_1^{(\alpha)}(\omega) &= \frac{2 (\alpha+1) \left(\alpha (\omega-2)^2-4 \omega+4\right)}{\omega^2}\,,\quad f_2^{(\alpha)}(\omega) = \left(\frac{4(\alpha+1) (\omega-1)}{\alpha \omega^2}\right)^{2 \alpha^2}\left(\frac{\omega}{\sqrt{1+\alpha}\ (\omega-2)}\right)^4\,.
\end{align*}

\subsubsection{Square case: Edelman's result}

In this section, we show how one recovers Edelman's results for the condition number in the \emph{square} matrix case $(N=M)$. Two of the results contained in Edelman's paper \cite{edel1} are as follows. For $\beta=1$,
\be
\lim_{N\to\infty}\mathcal{P}[\hat\kappa/N<x]=\mathrm{e}^{-2/x-2/x^2}, 
\ee
and for $\beta=2$
\be
\lim_{N\to\infty}\mathcal{P}[\hat\kappa/N<x]=\mathrm{e}^{-4/x^2}, 
\ee
where $\hat\kappa=\sqrt{\kappa}$ (as defined in our Letter). Thus, $\mathcal{P}[\hat\kappa<Nx]=\mathcal{P}[\kappa<N^2 x^2]$. Our result reads $\mathcal{P}[\kappa<x]\approx \exp\left(-\beta N^2 \Phi_-(x)\right)$. So all we have to do is to compute $\Phi_-(N^2 x^2)$ for $\alpha= 0$ to leading order in $N$. We get
\be
N^2\Phi_-(N^2 x^2)\sim \frac{2}{x^2},\quad\text{for }N\to\infty,
\ee
implying
\be
\mathcal{P}[\hat\kappa<Nx]\approx \exp\left(-\frac{2\beta}{x^2}\right),
\ee
which correctly reproduce Edelman's result for $\beta=2$ and his leading term in $N$ for $\beta=1$. The missing contribution in the latter case corresponds to a $\mathcal{O}(N)$ correction in the Coulomb gas approach not captured by our theory.

\subsection{Right rate function}

First, we give the definition of the functions $g^{(\alpha)}(x)$ and $h^{(\alpha)}(x)$ appearing in Eq. (13) of the Letter, as

\begin{align}
g^{(\alpha)}(x) &=\Xi\left(\sqrt{\varpi(x,t)+1},t\right),\\
h^{(\alpha)}(x) &=\frac{\sqrt{\varpi(x,t)}}{1+\sqrt{\varpi(x,t)+1}},\\
\varpi(y,\tau) &=\frac{4\tau(1+y)}{(\tau-1)^2 (y-\langle\kappa\rangle)},\\
\Xi(r,s) &=\frac{s^2 (r+1)^2-(r-1)^2}{s^2 (r-1)^2-(r+1)^2},
\end{align}
where $t=\sqrt{1+\alpha}$.\\
Next, we derive the difference in Coulomb fluid free energy when the minimum eigenvalue is pinned below the left edge $z_-$ of the MP law and the maximum eigenvalue above the right edge $z_+$ (at the positions $L$ and $U$ respectively). We want to estimate $\Delta E=E(L,\lambda_2,\ldots,\lambda_{N-1},U)-E_{\mathrm{mp}}(\lambda_1,\ldots,\lambda_N)$, where the initial energy $E_{\mathrm{mp}}$ is the one corresponding to the unperturbed MP configuration. Recall that the Coulomb energy is
\beeq{
E(\lambda_1,\ldots,\lambda_N)=\frac{1}{2}\sum_{i=1}^ N\lambda_{i}-\left(\frac{\beta}{2}(\alpha N+1)-1\right)\sum_{i=1}^N\ln \lambda_i-\frac{\beta}{2}\sum_{i\neq j}\ln|\lambda_i-\lambda_j|.
}
Then we have (after pinning the extreme particles)
\beeq{
E(L,\lambda_2,\cdots,\lambda_{N-1},U)&=\frac{L+U}{2}+\frac{1}{2}\sum_{i=2}^{N-1}\lambda_{i}-\left(\frac{\beta}{2}(\alpha N+1)-1\right)(\ln L+\ln U)-\left(\frac{\beta}{2}(\alpha N+1)-1\right)\sum_{i=2}^{N-1}\ln \lambda_i\\
&-\beta\sum_{2\leq i<j\leq N-1}\ln|\lambda_i-\lambda_j|-\beta\sum_{i=2}^{N-1}\ln|L-\lambda_i|+\beta\sum_{i=2}^{N-1}\ln|\lambda_i-U|-\beta\ln|U-L|.
}
Rescaling the eigenvalues $\lambda_i\to \beta N\lambda_i$, and recalling that $\lambda_{{\rm min}}\sim z_{-} $, $\lambda_{{\rm max}}\sim z_{+}$ we write the energy difference as $\Delta E(L,U)=\beta N\Delta e(L,U)$ with
\begin{equation}
\Delta e(L,U) =\frac{(L-z_-)+(U-z_+)}{2}-\frac{\alpha}{2}\ln\frac{LU}{z_- z_+}+\int_{z_-}^{z_+}\dd\eta \rho_{\mathrm{mp}}(\eta)\ln\left|\frac{(L-\eta)(\eta-U)}{(z_- - \eta)(\eta-z_+)}\right|,\label{deltaesupp}
\end{equation}
By replacing $L=\xi$ and $U=x\xi$,  and denoting the corresponding energy difference as $\Delta e(\xi,\xi x)$, we obtain Eq. \eqref{deltae1} of the Letter.\\
The integrals appearing in this expression can be evaluated analytically. We obtain the following results
\beeq{
\int_{z_-}^{z_+}\dd\eta \rho_{\mathrm{mp}}(\eta)\ln\left|\frac{\xi-\eta}{z_- - \eta}\right| &=\frac{1}{2} \Bigg(2(2+\alpha)\arcsinh\left(\frac{1}{2}\sqrt{\frac{z_--\xi}{\sqrt{1+\alpha}}}\right)+\alpha \Bigg(\ln \left(\frac{z_{-}}{\xi}\right)\\
&-2 \arctanh\left(\sqrt{\frac{z_+(z_--\xi)}{z_{-}(z_+-\xi)}}\right)\Bigg)-z_-+\xi+\sqrt{(z_--\xi) (z_+-\xi)}\Bigg)\,\\
\int_{z_-}^{z_+}\dd\eta \rho_{\mathrm{mp}}(\eta)\ln\left|\frac{\eta-\xi x}{\eta-z_+}\right|&=\frac{1}{2} \Bigg(\alpha  \left(\ln\frac{z_{+}}{\xi x}-2\arctanh\left(\sqrt{\frac{z_{-} (\xi x-z_+)}{z_{+}(\xi x-z_{-})}}\right)\right)\\
&+2(2+\alpha) \arcsinh\left(\frac{1}{2}\sqrt{\frac{\xi x-z_{+}}{\sqrt{1+\alpha}}}\right)+\xi x-z_{+}-\sqrt{(\xi x-z_{-})(\xi x-z_{+})}\Bigg).
}
After some algebra one can arrive to the intermediate expression
\beeq{
\Delta e(\xi,\xi x)&=-\frac{1}{2} \Bigg[-2\alpha  \left(\arctanh\left(\sqrt{\frac{z_+(z_--\xi)}{z_{-}(z_+-\xi)}}\right)+\arctanh\left(\sqrt{\frac{z_{-} (x\xi-z_+)}{z_{+}( x\xi-z_{-})}}\right)\right)\\
&+2(2+\alpha)\left( \arcsinh\left(\frac{1}{2}\sqrt{\frac{x\xi-z_{+}}{\sqrt{1+\alpha}}}\right)+\arcsinh\left(\frac{1}{2}\sqrt{\frac{z_--\xi}{\sqrt{1+\alpha}}}\right)\right)\\
&-\sqrt{(x\xi-z_{-})(x\xi-z_{+})}+\sqrt{(z_--\xi) (z_+-\xi)}\Bigg]\,.
}
Recall that the tail cumulative will be given by $\mathcal{P}[\kappa>x]\approx\exp\left[-\beta N \Phi_{+}(x)\right]$, where $\Phi_{+}(x)=\Delta e(\xi^{\star}(x),\xi^\star(x)x)$ so all is left to do is to minimize $\Delta e$ with respect to $\xi$. One can convince oneself that this is given by
\beeq{
\frac{\dd}{\dd \xi} \Delta e(\xi,\xi x) \Big|_{\xi=\xi^\star}=0\Rightarrow \xi^{\star}(x)=\frac{2(2+\alpha)}{x+1}
}
Plugging this result back to $\Delta e(\xi^{\star}(x),\xi^\star(x) x)$ and after some algebra we get to the result \eqref{rateright} of this Letter.

\subsection{Order of phase transitions}
An analysis of the derivatives of the rate functions will reveal the type of transition as they approach $\langle\kappa\rangle$. We obtain
\beeq{
\Phi_{-}(x)&=-\frac{(-1+\sqrt{1+\alpha})^8}{96\sqrt{1+\alpha}(1+\sqrt{1+\alpha})^4}(\langle \kappa\rangle-x)^{3}+\cdots\\
\Phi_{+}(x)&=\frac{\sqrt{2} \sqrt[4]{\alpha+1} \left(\sqrt{\alpha+1}-1\right)^4}{3 \sqrt{\alpha+2} \left(\sqrt{\alpha+1}+1\right)^2}(x-\langle \kappa\rangle)^{3/2}+\cdots
}

\subsection{Numerical simulations for the left rate function}
Let us recall the expression for the cumulative distribution
\beeq{
\mathcal{P}[\kappa<x]&=\frac{1}{(N-1)!}\int_0^\infty \dd\lambda_1 \left[\int_{\lambda_1}^{x\lambda_1} \dd \lambda_2\cdots \int_{\lambda_1}^{x\lambda_1}\dd\lambda_{N} P_\beta(\bm{\lambda})\right]
}
Suppose we were to implement a standard Monte Carlo simulation of the Coulomb fluid. After relaxation has been achieved, we  would simply use the Monte Carlo Markov Chain to estimate $\mathcal{P}[\kappa<x]$ by keeping a record of the number of states such that $\lambda_{i}\in[\lambda_1,x\lambda_1]$ for $i=2,\ldots,N$ divided by the total number of recorded states, that is
\beeq{
\mathcal{P}[\kappa<x]\propto \frac{\#(\{\lambda_1,\ldots,\lambda_N\}|\lambda_{i}\in[\lambda_1,x\lambda_1], i=2,\ldots,N )}{\#(\{\lambda_1,\ldots,\lambda_N\})}
}
There is nothing wrong with this procedure except that it is extremely inefficient: for  atypical values of $x$ it will be very unlikely to find the Coulomb fluid in that configuration, and a huge number of samples will be needed to have reliable statistics.\\
There is a smart way around this, though. One notices that it is very easy to obtain the samples $\#(\{\lambda_1,\ldots,\lambda_N\}|\lambda_{i}\in[\lambda_1,x\lambda_1], i=2,\ldots,N )$ by simulating the fluid directly with such constraints. The tricky part is to realize how to get the cumulative by using this subset of samples. We write 
\beeq{
\mathcal{P}[\kappa <y]&=\mathcal{P}[\kappa <y|y\leq x]K_{x}
}
where $K_x\equiv \mathcal{P}[y\leq x]$ depends only on $x$. The idea is then to simulate the Coulomb fluid such that $\lambda_{i}\in[\lambda_1,x\lambda_1]$ and to calculate the cumulative $\mathcal{P}[\kappa <y|y\leq x]$. Obviously the cumulative will be concentrate  mainly in the interval $y\in[x-\delta,x]$ for some small $\delta$. We can then use this to estimate the derivative of the left rate function, getting rid of the constant $K_x$, that is
\beeq{
\Phi'_{-}(y)=-\frac{1}{\beta N^2}\frac{\dd}{\dd y}\ln \mathcal{P}[\kappa <y|y\leq x]
}
Therefore, the algorithmic procedure is the following:
\begin{enumerate}
\item Choose a value of $x$.
\item Simulate the Coulomb fluid with barriers such that $\lambda_{i}\in[\lambda_1,x\lambda_1]$ and let it thermalize.
\item Construct the histogram of the cumulative $ \mathcal{P}[\kappa <y|y\leq x]$ of this confined fluid.
\item Estimate the derivative $\Phi'_{-}(y)$ using the constructed histogram
\end{enumerate}
Then by scanning the values of $x$ we obtain the derivative of the left rate function in the interval $x\in[1,\langle \kappa\rangle]$.

\end{widetext}

\end{document}